\documentclass[
 amsmath,amssymb,
 twocolumn,
 aps,
  prl,
  superscriptaddress,
float,
floatfix,
]{revtex4-1}

\usepackage{amsmath}
\usepackage{physics}
\usepackage{graphicx}
\usepackage{hyperref}
\usepackage{times}
\usepackage{xcolor}
\usepackage{multirow}
\usepackage{tabularx}
\usepackage{xcolor}

\newcommand{\subref}[2]{\hyperref[#1]{\ref*{#1}(#2)}}
\definecolor{refblue}{HTML}{2E2E91}
\hypersetup{colorlinks=true, citecolor=refblue, urlcolor=refblue, linkcolor=refblue}

\usepackage{graphicx}
\usepackage{dcolumn}
\usepackage{bm}
\usepackage{times}

\usepackage{filecontents}

\begin{document}

\preprint{APS/123-QED}

\title{High-fidelity imaging of a band insulator in a three-dimensional optical lattice clock}


\author{William R. Milner}
\email{william.milner@colorado.edu}
\affiliation{JILA, NIST and University of Colorado, 440 UCB, Boulder, Colorado 80309, USA}

\author{Lingfeng Yan}
\affiliation{JILA, NIST and University of Colorado, 440 UCB, Boulder, Colorado 80309, USA}

\author{Ross B. Hutson}
\affiliation{JILA, NIST and University of Colorado, 440 UCB, Boulder, Colorado 80309, USA}

\author{Christian Sanner}
\affiliation{JILA, NIST and University of Colorado, 440 UCB, Boulder, Colorado 80309, USA}

\author{Jun Ye}
\email{ye@jila.colorado.edu}
\affiliation{JILA, NIST and University of Colorado, 440 UCB, Boulder, Colorado 80309, USA}

\begin{abstract}
We report on the observation of a high-density, band insulating state in a three-dimensional optical lattice clock. Filled with a nuclear-spin polarized degenerate Fermi gas of $^{87}$Sr, the 3D lattice has one atom per site in the ground motional state, thus guarding against frequency shifts due to contact interactions. At this high density where the average distance between atoms is comparable to the probe wavelength, standard imaging techniques suffer from large systematic errors. To spatially probe frequency shifts in the clock and measure thermodynamic properties of this system, accurate imaging techniques at high optical depths are required. Using a combination of highly saturated fluorescence and absorption imaging, we confirm the density distribution in our 3D optical lattice in agreement with a single spin band insulating state. Combining our clock platform with this high filling fraction opens the door to studying new classes of long-lived, many-body states arising from dipolar interactions. 

\end{abstract}

\pacs{Valid PACS appear here}
\maketitle


\maketitle

Optical lattice clocks integrate quantum many-body physics and precision metrology to achieve state-of-the-art measurement precision \cite{bothwell2022resolving, mcgrew2018atomic, aeppli2022hamiltonian, bloom2014optical, nicholson2015systematic}. To advance clock performance, one wishes to probe as many atoms as feasible for the longest possible coherence time. Improvements in both precision and accuracy of optical lattice clocks, with increased atom numbers, have been enabled by the development of high-fidelity, microscopic imaging of the atomic cloud to spatially resolve clock shifts~\cite{marti2018imaging, campbell2017fermi}. The combination of high density and long coherence time will allow characterization of novel systematic effects such as that arising from dipolar interactions between atoms on neighboring lattice sites~\cite{chang2004controlling, cidrim2021dipole, kramer2016optimized, rossdipolar}. Lattice thermometry \cite{hofrichter2016direct} and studies of novel physics such as SU($N$) magnetism \cite{gorshkov2010two, taie20126} will also benefit from accurate imaging at high density where these phenomena emerge.

To optimize atom number while minimizing interaction-related dephasing, a clock platform based on a 3D lattice geometry and Fermi-degenerate matter has been developed \cite{campbell2017fermi, hutson2019engineering}.  Following nuclear spin polarization \cite{sonderhouse2020thermodynamics, stellmer2011detection}, the Pauli exclusion principle mandates there is at most one atom per lattice site in the ground motional state.  To ensure this ground state motional occupation during lattice loading we operate with $k_{B}T < k_{B}T_{F} < \hbar\omega_{bg}$, where $T$,  $T_{F}$, $ \hbar\omega_{bg}$ refers to the atomic temperature, Fermi temperature and lattice bandgap respectively \cite{will2012atom}. At the highest density affordable with one fermion per lattice site, this system realizes an insulating state of matter where tunnelling is suppressed~\cite{hutson2019engineering, schneider2008metallic}. Combining this high-density system with spin-orbit coupling generated from clock addressing will enable exploring cluster state generation and tunable spin models \cite{mamaev2019cluster, mamaev2021tunable}.


Differential frequency shifts across the optical lattice encoding potential systematic effects can be spatially resolved  by combining \textit{in situ} imaging and narrow-line clock spectroscopy \cite{marti2018imaging}.  To extract these frequency shifts, two subsequent images of the ground and excited state density distributions are required. Thus for our clock platform, accurate \textit{in situ} imaging at high density is imperative. In our lattice where the average distance between atoms ($407$ nm) is comparable to the probe wavelength ($461$ nm), imaging with a weak, resonant probe is strongly perturbed.  Both collective effects mediated by dipolar interactions \cite{andreoli2021maximum} and systematic defects such as lensing of the probe beam \cite{ketterle1999making, rath2010equilibrium} introduce errors to the reconstructed density distribution at high density. 

To mitigate these systematic effects, different techniques can be used to reduce the absorption cross section and make the cloud "optically thin".  These techniques can be broadly divided into two categories: dispersive imaging at large detuning from resonance \cite{kadlecek2001nondestructive, bradley1997bose, andrews1996direct} and saturated imaging at high intensity \cite{depue2000transient, reinaudi2007strong, yefsah2011exploring, chomaz2012absorption}. For dispersive imaging extracting information about the atomic density often requires spatially filtering the scattered and unscattered light in the Fourier plane of the imaging system, demanding precise fabrication and alignment of custom optics.  Additionally, careful studies of dispersive imaging show that residual systematic effects at finite detuning are non-negligible and can be addressed using differential measurement schemes at opposite detuning \cite{lee2012compressibility}. To address these imaging errors in this work, we use both highly saturated fluorescence and absorption imaging.

In this Letter, we report on the observation of a band insulating state in our 3D optical lattice clock. Using highly saturated imaging to mitigate imaging errors, with a saturation parameter far greater than the optical depth, we accurately confirm the density distribution in our 3D optical lattice in good agreement with thermodynamic calculation. We extend previous work using high intensity fluorescence imaging \cite{depue2000transient},  confirming the accuracy of this imaging technique in a new high density regime with a degenerate Fermi gas of $^{87}$Sr \cite{sonderhouse2020thermodynamics, sanner2021pauli}.  With atomic densities as high as $6 \times 10^{14}$ atoms/cm$^{3}$, we observe systematic agreement with atom counts obtained via time-of-flight absorption imaging and identify the range where the extracted atomic density distribution is not blurred by our imaging pulse.  

\begin{figure}[t]
    \centering
    \includegraphics[width=8cm]{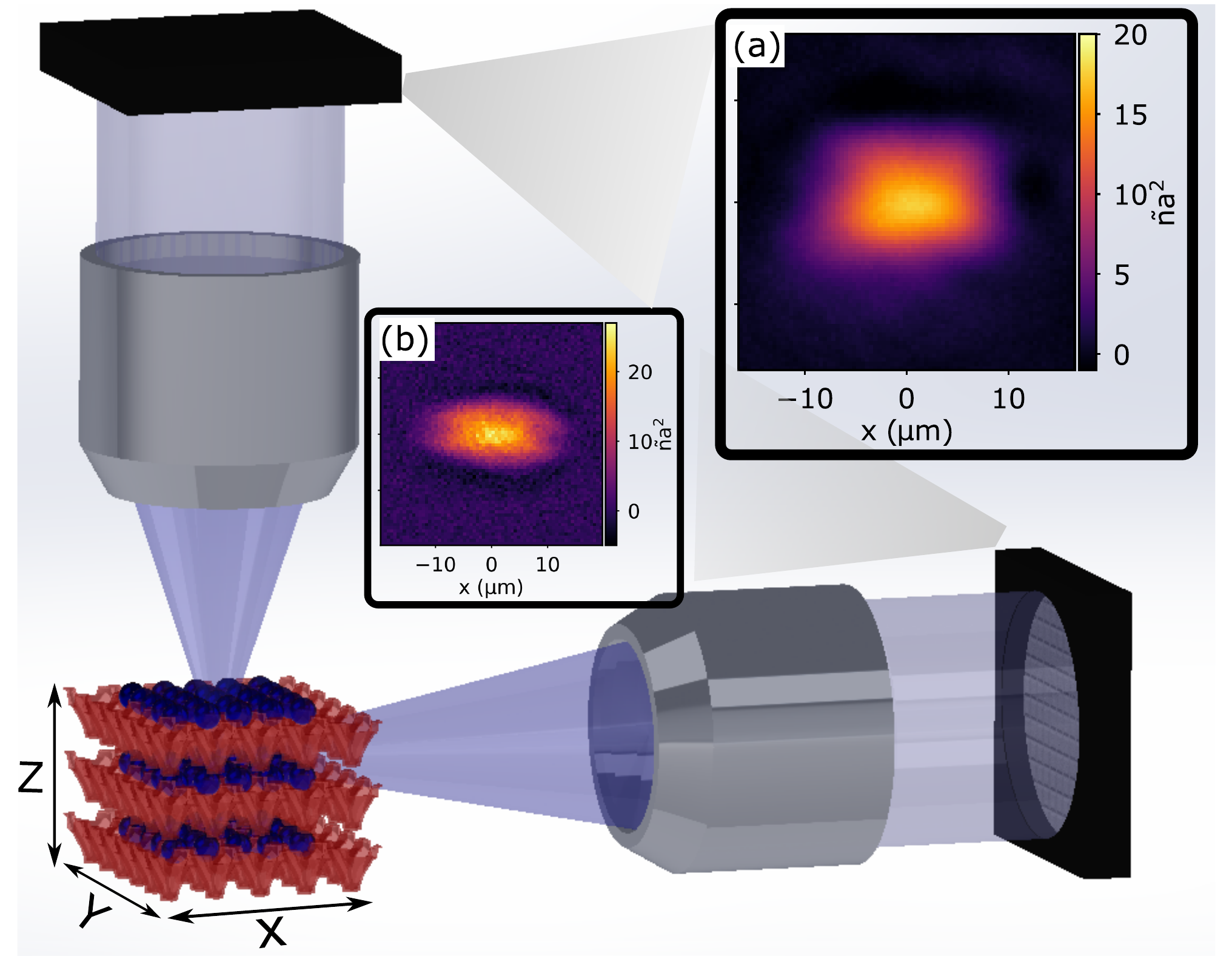}
    \caption{Schematic of our clock platform. Vertical and horizontal imaging systems with numerical apertures of $0.2$ and $0.1$ respectively provide measurements of the 2D density distribution $\tilde{n}$.  Accounting for the lattice spacing $a$ $=$ $407$ nm,  $\Tilde{n} a^{2}$ is determined from highly saturated absorption imaging. To mitigate imaging errors, the atoms are highly saturated and each scatters photons with a maximum rate of $\Gamma$/2. Measurements from our high resolution imaging system integrated along gravity are presented in panel (a), where the density distribution is extracted for thermodynamic modeling. Images from the horizontal imaging system in panel (b) are just used to determine our atom cloud aspect ratio for our inverse Abel transform.}
    \label{fig:fig1}
\end{figure}

Our high intensity imaging scheme is outlined in Fig.~\ref{fig:fig1}. The combination of atomic level structure and relatively large mass of $^{87}$Sr is particularly well suited for our imaging technique, providing a cycling transition with a large scattering rate while avoiding significant motional effects from the imaging pulse. The transition from $^{1}S_{0}$ to $^{1}P_{1}$ with linewidth $\Gamma = 2 \pi \times 30.5$ MHz provides a large photon scattering rate with minimal depumping to dark states during the imaging time \cite{barker2015enhanced}. During a 1 $\mu$s pulse at full saturation about $100$ photons per atom are scattered and the atoms accelerate at $a = \frac{\hbar k \Gamma}{2 m}$ where $k$ is the
imaging light wavenumber and $m$ is the atomic mass. The net momentum transfer amounts to a Doppler shift of $k a \tau$ = 2.8 MHz which is much less than the transition linewidth $\Gamma$/$2 \pi$. Finally, the linear displacement for a $1$ $\mu$s pulse at full saturation is just $\frac{a \tau^{2}}{2}$ =  $0.6$ $\mu$m. This linear displacement and corresponding Doppler shift can be largely cancelled in fluorescence imaging by retroreflecting the incident beam.  The spread in transverse position due to random momentum transfer from spontaneous emission is $\frac{\hbar k}{6 m}t^{3/2}\sqrt{\Gamma / 2}$   $ < 0.1$ $\mu$m over a 1 $\mu$s pulse duration and small compared to our $1.3$ $\mu$m imaging resolution \cite{joffe1993transverse}. Using highly saturated absorption imaging, we measure the column density distribution $\tilde{n}$ in our optical lattice in Fig.~\subref{fig:fig1}{a}.  Accounting for the lattice spacing $a$ $=$ $407$ nm corresponding to the $^{87}$Sr magic wavelength at $813$ nm,  the scaled column density $\Tilde{n} a^{2}$ is plotted.

Saturated absorption and fluorescence imaging are beneficial in comparison to standard imaging techniques in a number of ways. In this highly saturated regime the scattering rate is largely immune to beam intensity, frequency, and pointing fluctuations.  Given the saturation intensity $I_{sat}$ =  40 mW/cm$^{2}$ for the imaging transition, a Gaussian probe beam with 20 mW of optical power and a 100 $\mu$m waist corresponds to a peak intensity of $I \sim 3000$ I$_{sat}$, within the typical constraints of a standard imaging laser system. Given that the probe beam is attenuated through the atom cloud, a saturation parameter $I/I_{sat}$ much greater than the optical depth is required to fully saturate the imaging transition. We note parallels between fluorescence and absorption imaging at high saturation. In both cases, the extracted atom number is determined by a single variable. For fluorescence imaging, this corresponds to the number of collected photons per atom and for saturated absorption imaging the number of missing photons per atom in the probe beam. Thus, both fluorescence and saturated absorption imaging can be calibrated via a single absolute atom number measurement. For images in our 3D lattice, we determine our atom number via clock excitation fraction fluctuations arising from quantum projection noise (QPN) \cite{itano1993quantum, Supplement}.   

For fluorescence imaging, only a single image of collected fluorescence in an arbitrary direction is required, minimizing fringing and simplifying image processing substantially. Fluorescence imaging also avoids limited dynamic range issues suffered from high intensity absorption imaging.  Strategies such as multiple measurements at varying intensity to determine the atomic density in different regions of the cloud may be taken to confront this issue \cite{yefsah2011exploring, chomaz2012absorption}. The primary disadvantage of fluorescence imaging in comparison to absorption imaging is that the signal-to-noise is generally worse \cite{Supplement}. To optimize signal-to-noise ratio (SNR) in fluorescence imaging, the photon collection efficiency and therefore the numerical aperture (NA) of the imaging system, must be maximized. In our experiment, the vertical and horizontal imaging systems have numerical apertures of $0.2$ and $0.1$, corresponding to collection efficiencies of approximately $1$ and $0.2$ percent. Alternatively, if spatial resolution is not required then the pulse duration can be extended enhancing the number of detected photons.

\begin{figure}
    \centering
    \includegraphics[width=8cm]{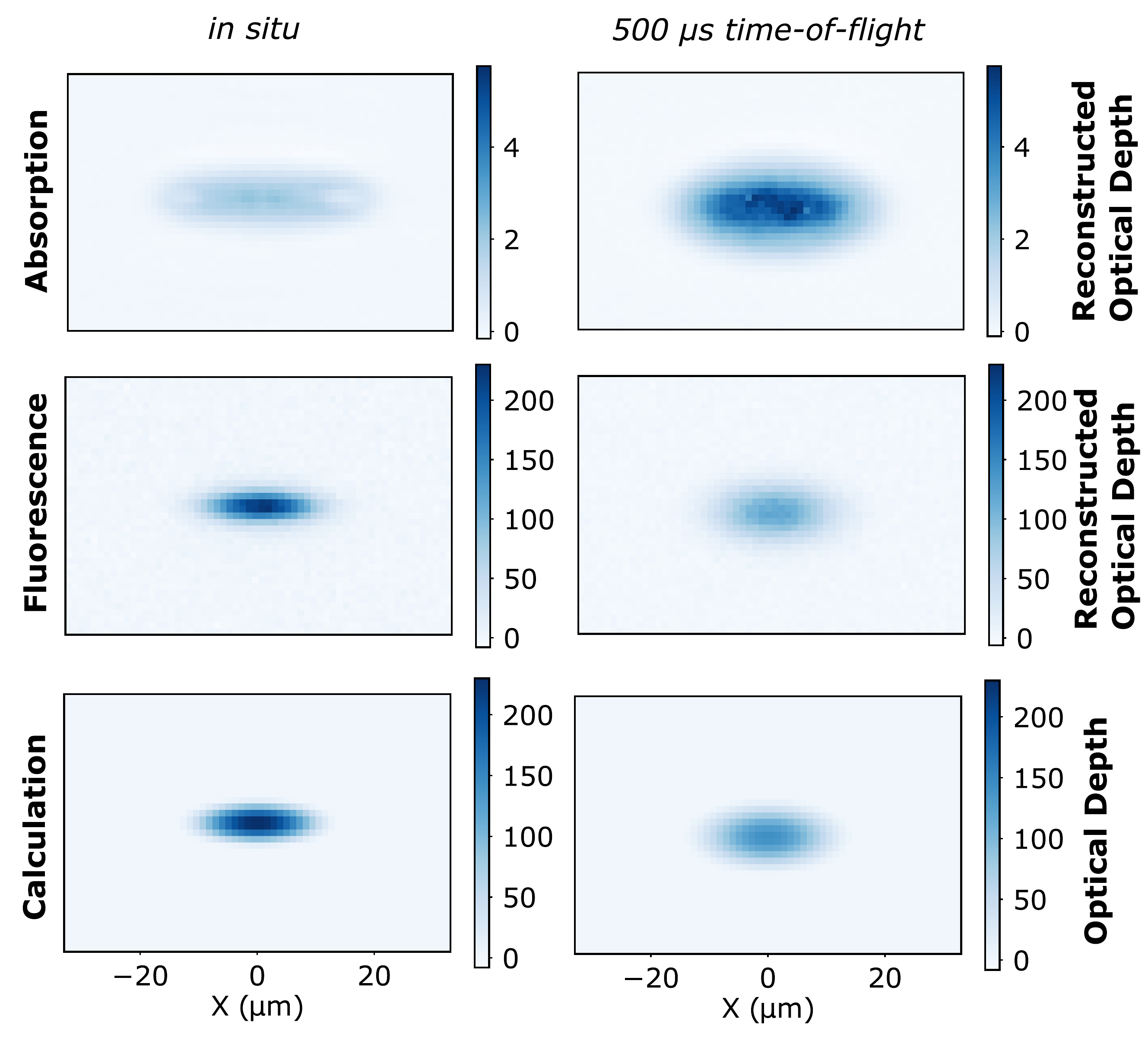}
    \caption{A comparison of high intensity fluorescence and standard absorption imaging ($I \sim I_{sat}$) at optical depths exceeding $200$ in our highly degenerate Fermi gas is shown. \textit{In situ} absorption imaging at low intensity yields strikingly erroneous measurements at high density. The calculated 2D Fermi gas distribution according to our experimental parameters is shared for comparison in qualitative agreement.   }
    \label{fig:fig2}
\end{figure}

To motivate the development of our high intensity imaging technique, systematic errors associated with standard \textit{in situ} imaging techniques at high density are presented in Fig.~\ref{fig:fig2}. Absorption imaging at I $\sim $ I$_{sat}$ and high intensity fluorescence imaging are presented side-by-side for comparison.  To study these systematic errors at high density, we prepare a sample with optical depth $> 200$ by producing a degenerate Fermi gas with 10 nuclear spin components, $\approx 2 \times 10^{5}$ atoms and a T/T$_{F}$ of approximately 0.1 in a crossed dipole trap. The errors associated with low intensity absorption imaging can be seen twofold. First, the reconstructed optical depth from absorption detection in the upper left panel is far too low, two orders of magnitude less than the expected value of $\sim 200$. This erroneously low optical depth is attributed to effects such as enhanced forward emission and lensing of probe light \cite{ketterle1999making}. Secondly, the reconstructed optical depth in the upper right panel increases after a $500$ $\mu$s time-of-flight expansion conclusively demonstrating the density dependence of these observed systematic errors. 

 In comparison, saturated fluorescence imaging yields a far larger reconstructed optical depth and diffuses following expansion as expected. We compare this reconstructed 2D density distribution with the expected distribution corresponding to a Fermi gas. Using independently measured experimental values,  we calculate this distribution with no free parameters \cite{zweirlein2008making}. The total atom number and reduced temperature T/T$_{F}$ are determined from time-of-flight absorption imaging at low density with an optical density $\sim 1$.  The trapping frequencies are extracted from parametric confinement modulation. Using these parameters, we calculate both an  \textit{in situ} and 500 $\mu$s time-of-flight Fermi gas profile for comparison with our measurements. We observe qualitative agreement between measurement and calculation at these extremely high optical depths.

\begin{figure}[h]
    \centering
    \includegraphics[width=8 cm]{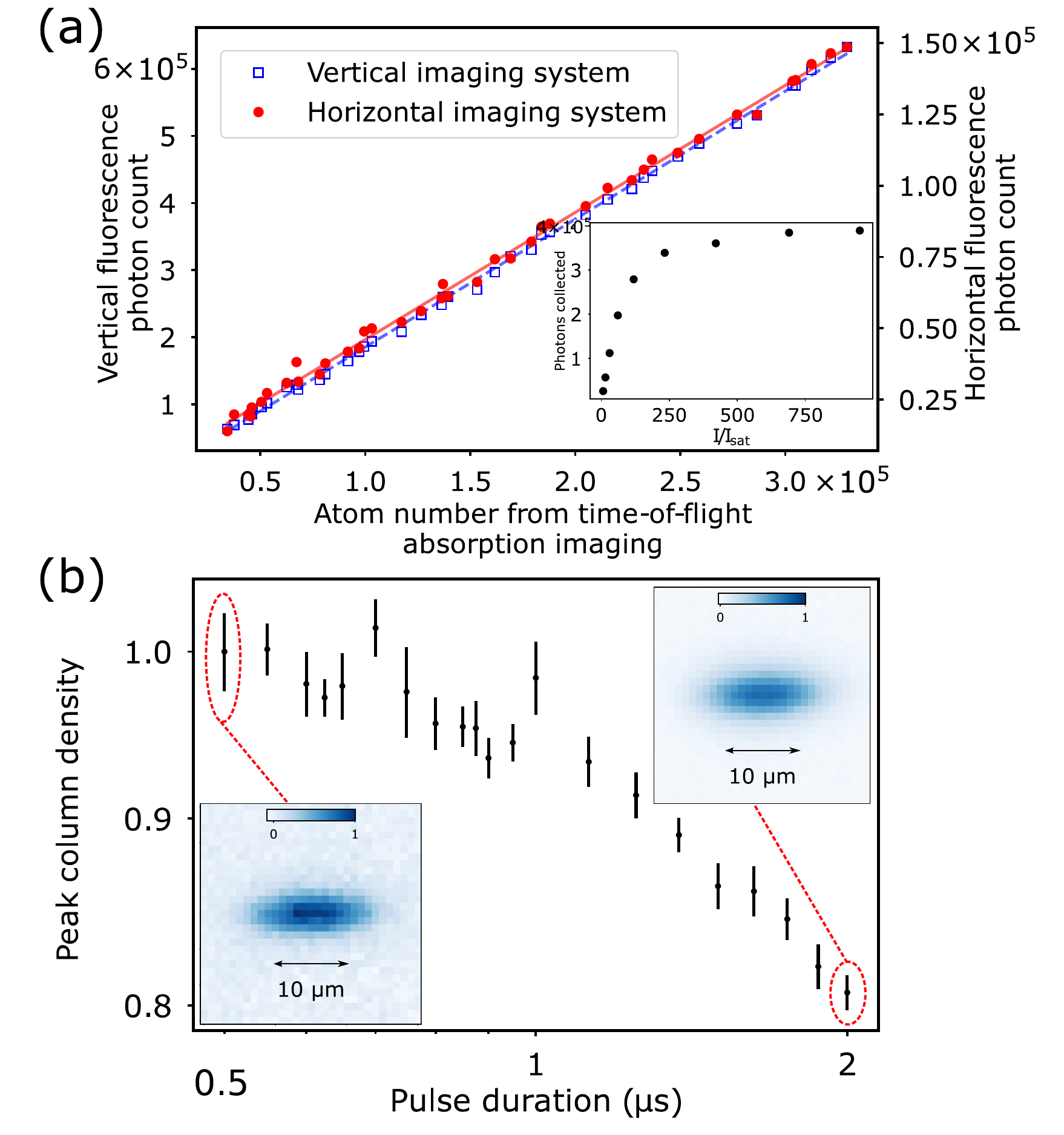}
    \caption{(a) Calibration method for \textit{in situ} fluorescence detection using atom counts from time-of-flight absorption imaging. Collected photon counts from both the vertical and horizontal imaging systems are plotted, with solid and dashed lines representing fits to the horizontal and vertical measurements respectively. Inset: Collected photon count with vertical imaging system as a function of I/I$_{sat}$ at 1 $\mu$s pulse duration. (b) Peak column density as a function of fluorescence pulse duration. Measurements are normalized by 1.9$\times 10^{11}$ atoms/$\textrm{cm}^{2}$, the column density at the shortest pulse duration of $500$ ns. Images at $500$ ns and $2$ $\mu$s in inset are plotted for comparison. The error bars denote the standard error of the mean.}
    \label{fig:fig3}
\end{figure}

Intrigued by the measurements presented in Fig.~\ref{fig:fig2}, we undertake a quantitative study on the fidelity of our saturated imaging technique. We present a calibration method for fluorescence detection, using the total number of collected fluorescence photons for comparison with an accurate atom number reference. Absorption imaging at low density following time-of-flight expansion serves as an appropriate calibration. Following expansion for 7 ms, the optical depth is $\sim 1$ and systematic imaging errors can be safely ignored.  To  independently calibrate the atom number in our 10 spin Fermi gas, we prepare a thermal sample and use measured density fluctuations to determine the effective absorption cross section \cite{sanner2010suppression, tobias2020thermalization, muller2010local}. In Fig.~\subref{fig:fig3}{a} we ensure this calibration shows systematic agreement with atom numbers between approximately $1 \times 10^{5}$ and $4 \times 10^{5}$, varied by increasing our final evaporation trap depth.  For the $3$ $\mu$s pulse duration used, the fitted calibration is in reasonable agreement with calculation using the measured quantum efficiency and imaging system numerical aperture \cite{Supplement}. To ensure that the imaging transition is fully saturated, the laser intensity at $1$ $\mu$s pulse duration is increased until the collected photon number plateaus, as seen in the figure inset. 

\begin{figure}[h]
    \centering
    \includegraphics[width=8cm]{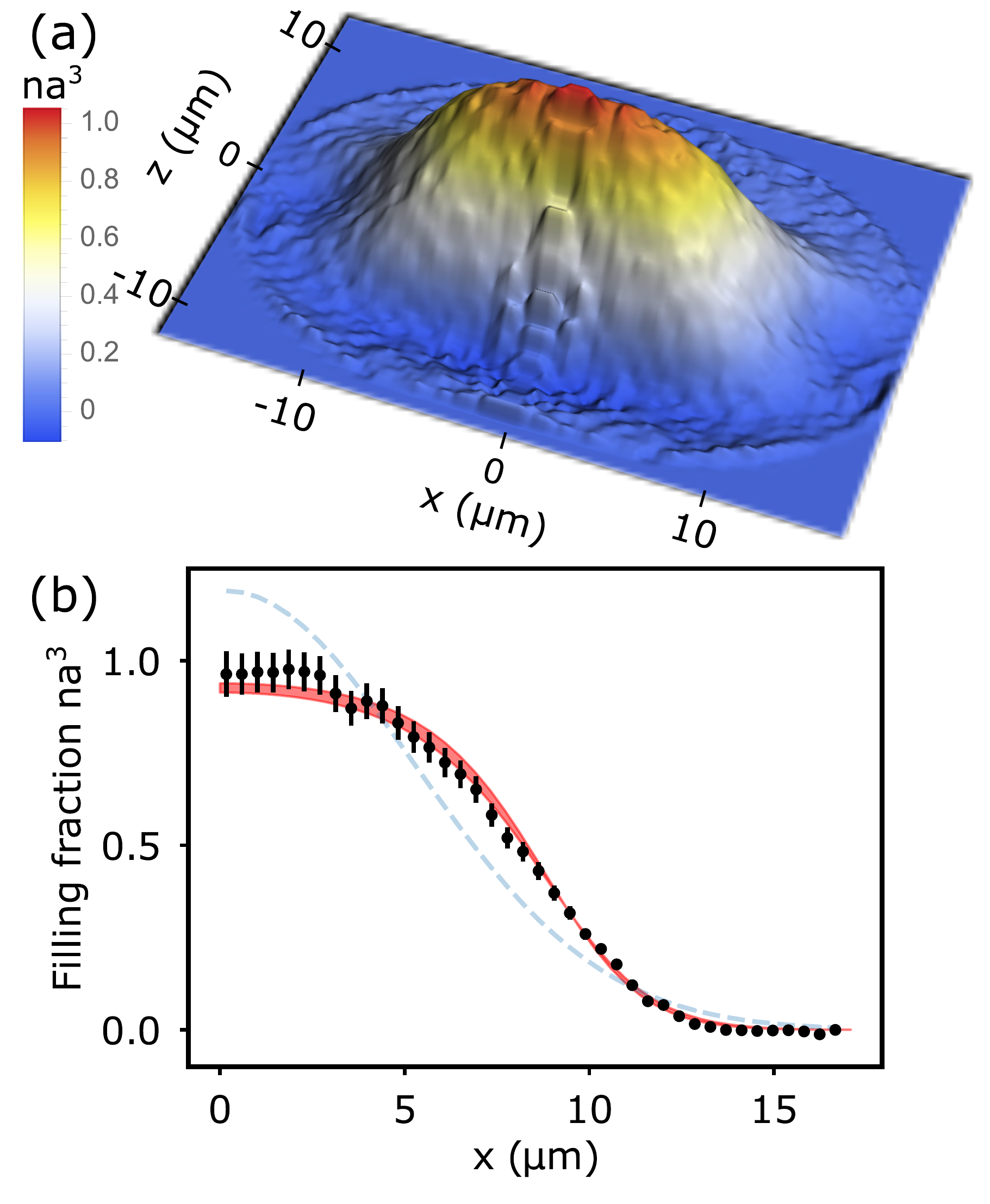}
    \caption{(a) The three-dimensional density distribution and the corresponding lattice filling fraction are determined from \textit{in situ} absorption image in Fig.~\subref{fig:fig1}{a} and the use of an inverse Abel transformation. (b) A linecut along $z = 0$  provides the data points in circle. Errorbars are both the statistical uncertainty of the Abel transformation and atom number uncertainty added in quadrature. We start with a prediction based on HTSE calculation, using independently measured values for the temperature, atom number, and harmonic confinement. The best fit to the data results in a $10\%$ reduction of the measured aspect ratio $\omega_{y}$/$\omega_{x}$ and $5\%$ reduction of the measured $T/T_{F}$. The red line captures this fit, with temperature uncertainty in the shaded band. The blue dashed line is a fit to Gaussian in qualitative disagreement with $n a^{3}$.}
    \label{fig:fig4}
\end{figure}

To perform accurate spatially resolved measurements, we must also determine the blurring induced by our imaging pulse. Just as collective effects introduce errors to the reconstructed density distribution, any systematic changes to $\Tilde{n}$ introduced by our imaging pulse must be determined. To calibrate this blurring in Fig.~\subref{fig:fig3}{b}, we extend the fluorescence pulse duration and examine the peak column density as atoms diffuse. The inset shows averaged images from 500 ns and 2 $\mu$s pulse durations. We note that we observe no atom loss or molecular formation over the full 2 $\mu$s range, confirmed by the detected photon count increasing linearly with pulse duration. To minimize blurring, we carefully retroreflect our probe beam by optimizing the backcoupled light through the probe optical fiber. At pulse durations up to 1 $\mu$s, we confirm that the peak column density decreases by $< 5 \%$. \cite{Supplement}.

Motivated by the calibration reported in Fig.~\ref{fig:fig3}, we directly determine the 3D density distribution in a deep optical lattice via saturated $\textit{in situ}$ absorption imaging. We form a cubic lattice with trap depths of approximately $60$, $70$, and $50$ $E_{r}$ in three orthogonal directions, where $E_{r}$ is the lattice photon recoil energy $\approx$ $h$ $\times$ $3.5$ kHz. Following forced evaporation with $10$ nuclear spin states we spin polarize using a focused beam detuned from the $^{3}P_{1}$ intercombination line to form a state-dependent potential, removing nearly all but the $m_{F}$ = -$9$/$2$ atoms \cite{sonderhouse2020thermodynamics, stellmer2011detection}. Clock spectroscopy confirms $\approx$ $90 \%$ spin purity. An additional step of spin purification is applied by coherently driving the $m_{F}$ = -$9$/$2$ atoms into the excited clock state and removing any residual spins with a resonant imaging pulse. Absorption imaging directly provides us with the column density distribution $\tilde{n}$, integrated through the vertical axis along gravity as depicted in Fig.~\subref{fig:fig1}{a}. Based on our Fig.~\subref{fig:fig3}{b} analysis, we choose a pulse duration of $1$ $\mu$s to minimize blurring and a saturation intensity of $54(4)$, substantially larger than peak optical density of $\sim 15$. To spatially probe the band insulator plateau we use an imaging magnification of $38.8$ to achieve an effective pixel size of $412$ nm, roughly equal to the lattice constant $a$ $=$ $407$ nm.  We note that our effective pixel size is smaller than our optical resolution of $1.3$ $\mu$m, thus our imaging system is optically oversampled. To extract the 3D density distribution, we use an inverse Abel transform \cite{dribinski2002reconstruction}.  Given our vertical imaging is not along an axis of cylindrical symmetry, $n$ must be appropriately scaled by the aspect ratio of the spatial density distribution \cite{Supplement}.  The aspect ratio is independently calibrated using the absorption imaging measurement in Fig.~\subref{fig:fig1}{b}.

At this high magnification, the SNR in fluorescence imaging for a 1 $\mu s$ pulse duration is limited by a combination of read noise and photon shot noise. We found that even after extensive averaging the extracted 3D density distribution using an inverse Abel transform was sensitive to small fluctuations in $\tilde{n}$.  Thus, saturated absorption imaging with a superior SNR provides a more robust technique to characterize the 3D density distribution. This extracted 3D density distribution is plotted in Fig.~\subref{fig:fig4}{a}.

To judge the fidelity of our measured 3D density distribution, we compare the line cut at $z = 0$ with calculation in Fig.~\subref{fig:fig4}{b}. To estimate the density distribution, we use a High Temperature Series Expansion (HTSE) calculation in the atomic limit \cite{hofrichter2016direct, Supplement, taie20126, hazzard2012high}. The ingredients of this calculation include values for the atomic temperature, harmonic confinement, and total atom number.  Given the density distribution only depends on the ratio of the respective harmonic confinements, the measured aspect ratios from Fig.~\ref{fig:fig1} are used for our HTSE calculation.  The total atom number $N$ is determined from quantum projection noise measurements \cite{Supplement}.  To estimate the temperature including heating during lattice loading, we measure the reduced temperature $T/T_{F}$ in time-of-flight after a round-trip from the lattice back to the dipole trap and determine an entropy-per-particle increase of $0.25(6)$ $k_{B}$. Inferring an entropy increase of $0.13(3)$ $k_{B}$ in a single lattice loading sequence, we estimate a $T/T_{F}$ of $0.165(7)$. Although we did not perform a cross-dimensional thermalization measurement to directly verify thermal equilibrium, the uncertainty in our temperature is included in the shaded band of the HTSE calculation in Fig.~\subref{fig:fig4}{b} \cite{monroe1993measurement, valtolina2020dipolar}.  We note that the extended plateau region is larger than our $1.3$ $\mu$m imaging resolution. To further quantify the imaging fidelity, we compare $n a^{3}$ to a Gaussian fit in clear disagreement with data.

In conclusion, we report on the observation of a spin-polarized, band insulating state in our 3D optical lattice clock. This has been enabled by characterizing saturated $\textit{in situ}$ imaging techniques to accurately determine our density distribution. Broadly, the saturated imaging techniques in this work will be applicable for studies of SU($N$) magnetism and thermodynamics in the Mott-insulating regime \cite{goban2018emergence, zhang2014spectroscopic}. 
With the high filling fraction demonstrated in this work, many-body states arising from dipolar interactions can be generated between atoms on neighboring lattice sites \cite{chang2004controlling, cidrim2021dipole}.

$\bold{Acknowledgement}$. We thank D. Kedar for maintaining the ultrastable clock laser used in this work and A. Aeppli, K. Kim, J. M. Robinson, M. Miklos, and Y. M. Tso for useful discussions. We thank K. Kim, N. D. Oppong, and L. Sonderhouse for careful reading of the manuscript and for providing insightful comments. Funding for this work is provided by NSF QLCI OMA-2016244, DOE Center of Quantum System Accelerator, DARPA, AFOSR, V. Bush Fellowship, NIST, and NSF Phys-1734006.

\bibliography{references}

\end{document}


\preprint{APS/123-QED}

\title{Supplemental material to \\High-fidelity imaging of a band insulator in a three-dimensional optical lattice clock}

\pacs{Valid PACS appear here}

\maketitle

\subsection{Density diffusion}
Here we provide supplemental analysis to the data presented in Fig.~3(b). In panel A of Fig.~\ref{fig:figS1}, we plot the integrated counts along the $x$ axis of each image. We see an asymmetry emerge along the direction of the probe beam as the pulse duration is extended. This asymmetry suggests that the observed density diffusion may arise from  inhomogeniety between the incident and retroreflected beams. While the power is certainly mismatched, this could also be due to either imperfect spatial alignment or mode mismatch given the divergence of the probe beam. 

We also plot the total counts in each image as a function of pulse duration in panel B. The linear character of the counts over the full pulse duration range suggests that we do not observe appreciable atom loss or pumping to dark states. The counts at each pulse duration are normalized to the counts at $500$ ns. The inset shows the Gaussian RMS width of the cloud as a function of pulse duration.

\renewcommand{\thefigure}{S1}
\begin{figure}
    \centering
    \includegraphics[width=7 cm]{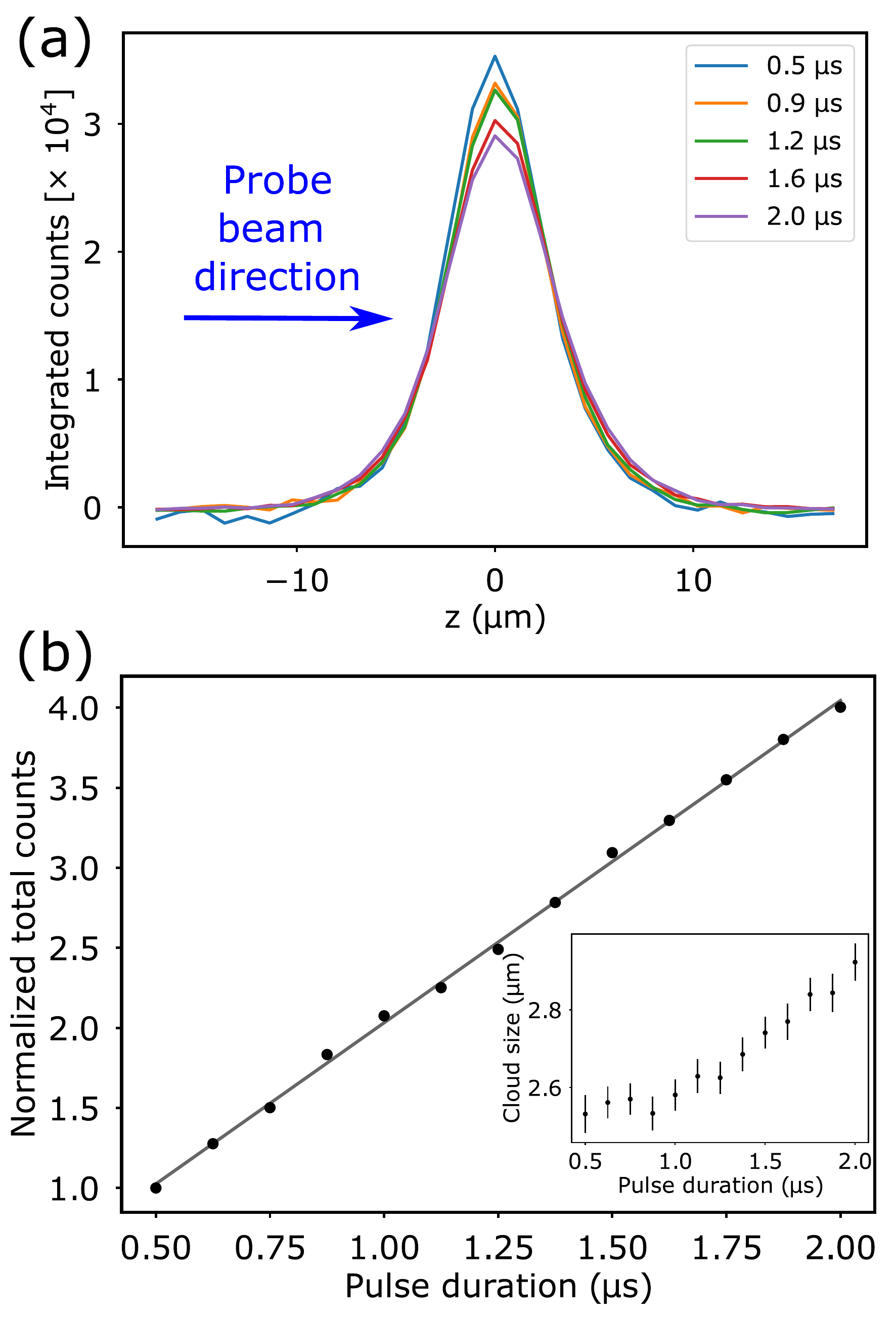}
    \caption{Panel (a) shows the integrated counts from the images in Fig.~3(b) of the main text along the $x$ axis as a function of pulse duration. The total counts at each pulse duration is plotted in panel (b), normalized by the counts at $500$ ns. Given the detected photon count increases linearly with pulse duration, we observe minimal atom loss or molecular formation over the full 2 $\mu$s range. The inset shows the Gaussian RMS width of the cloud as a function of pulse duration.}
    \label{fig:figS1}
\end{figure}

\subsection{Signal-to-noise comparison}
In the main text of the paper we refer to both saturated absorption and fluorescence imaging. We provide a quantitative comparison of the signal-to-noise ratio (SNR) between the two techniques here. We express our signal-to-noise for a detection pixel in terms of the normalized variance $\mathbb{V}(N)/N$, where $N$ denotes the number of atoms within the respective detection region.  For fluorescence imaging the SNR is simply determined by the shot noise associated with the number of detected photons. To calculate the total atom number, we first convert the fluorescence counts detected on our camera to the number of collected photons. Then, using the collection efficiency of our imaging system and scattering rate of our atomic transition we determine the conversion of detected photons per atom. On our CCD camera, we measure $n_{a}$ counts in a given pixel. Using the quantum efficiency $q$ of the imaging system, and the camera conversion gain $g$ in units of counts per photo electron, we infer $\frac{n_{a}}{qg}$ photons.  At full saturation, the atomic scattering rate is $\frac{\Gamma}{2}$ and the number of photons scattered per atom is $P_{sc} = \frac{\Gamma}{2} \times \tau$, where $\tau$ is the pulse duration. Finally, we denote the collection efficiency as $Y$, determined by the numerical aperture of our imaging system and by radiation pattern anisotropies.  Combining terms, the total atom number is $N  = \frac{n_{a}}{gqYP_{sc}}$. Using error propagation, we determine the variance $\mathbb{V}_{Fl}(N)$.

\begin{equation} \mathbb{V}_{Fl}(N)  = \Big( \frac{\partial N}{\partial n_{a}} \Big)^2\mathbb{V}(n_{a}) = \Big( \frac{1}{gqYP_{sc}} \Big)^{2}gn_{a} \end{equation} 

Here, we have used the fact that the distribution of generated photo electrons $n_{e}$ is binomial. Thus, $\mathbb{V}(n_{a}) = \mathbb{V}(g \times n_{e}) = g^{2}\mathbb{V}(n_{e}) = g^{2}n_{e} = g n_{a}$. Combining terms: \begin{equation} \mathbb{V}_{Fl}(N)/N = \frac{1}{qYP_{sc}}\end{equation} 

The SNR associated with absorption imaging is more complicated given the formula for the atom number in Eq. 3 has both logarithmic and linear terms and involves two images $n_{a}$ and $n_{b}$ with and without atoms present. Here, $A$ and $\sigma_{0}$ refer to the effective pixel size accounting for the imaging system magnification and effective atomic absorption cross section, respectively. Similar to fluorescence imaging, an appropriate error propagation of the $n_{a}$ and $n_{b}$ terms determines Eq. 4 and Eq. 5. We summarize the formulas here and point a reader to reference \cite{marti2014scalar} for a full derivation.

\begin{equation}  N = \frac{A}{\sigma_{0}} \textrm{log}(\frac{n_{b}}{n_{a}}) + \frac{2}{\Gamma \tau g q} (n_{b} - n_{a})\end{equation}  

 \begin{equation} \mathbb{V}_{Abs}(N) = g \tilde{A}^{2} (\frac{1}{n_{a}} + \frac{1}{n_{b}}) + g \tilde{B}^{2} (n_{a} + n_{b}) + 4g\tilde{A}\tilde{B}  \end{equation}

 \begin{equation} \tilde{A} = \frac{A}{\sigma_{0}}, \tilde{B} = \frac{2}{qg \tau \Gamma}  \end{equation}


We compare the different techniques in Fig.~\ref{fig:figS2} using the experimentally relevant parameters for our imaging system. In both cases, a $1$ $\mu$s resonant pulse is used with a numerical aperture of $0.2$ and a quantum efficiency of $85 \%$. For the fluorescence SNR in blue, the transition is assumed to be fully saturated and scatters photons with a rate of $\Gamma/2$. For the $I/I_{sat}$ = $\sim 55$  we use for our inverse Abel measurements, the SNR in absorption imaging is superior to fluorescence imaging in regions where the column density is higher than $2$ atoms/$a^{2}$. Particularly given our peak density of $\tilde{n}a^{2}$ = $\sim 20$ in Fig.~1(a), absorption imaging provides a better SNR in the regions of high density where we extract our peak filling fraction. At a critical OD of $0.17$, fluorescence detection under our experimental parameters provides a superior SNR at all imaging intensities. We note these calculations neglect technical noise, in particular camera readout noise, which can be accounted for by offsetting $\mathbb{V}(n_{a})$ accordingly. This contribution will disproportionately reduce the SNR of fluorescence imaging, as the fluorescence counts are substantially lower than the absorption counts.

To probe fine spatial details in our atomic cloud, an imaging resolution smaller than the length scale of these spatial features is required. To achieve this condition, a sufficiently large numerical aperture imaging system must be utilized and aberrations must be minimized. In this case, the imaging resolution is fundamentally limited by diffraction. We verified the diffraction-limited performance of our $NA$ = $0.2$ objective lens by propagating a point source at $461$ nm through a test setup (including all imaging path optics and vacuum viewports) and measuring the point-spread function.

While absorption and fluorescence imaging rely on the same light scattering process (they only collect different parts of the scattered EM field \cite{ketterle1999making}), the signal amplitudes for these two methods scale differently with the $NA$. When collecting fluorescence, the solid angle coverage of the imaging system proportionally affects the signal down to the lowest spatial frequencies. This is not the case for absorption imaging, where the amplitude of spatial frequency components below the $NA$-dependent bandwidth is constant as the $NA$ is further increased (assuming the lens fully covers the probe beam). In other words, for fluorescence imaging, most of the signal light gets collected in the outer ring fraction of the lens aperture, which renders it particularly susceptible to lens imperfections.

\renewcommand{\thefigure}{S2}
\begin{figure}[t]
    \centering
    \includegraphics[width=8 cm]{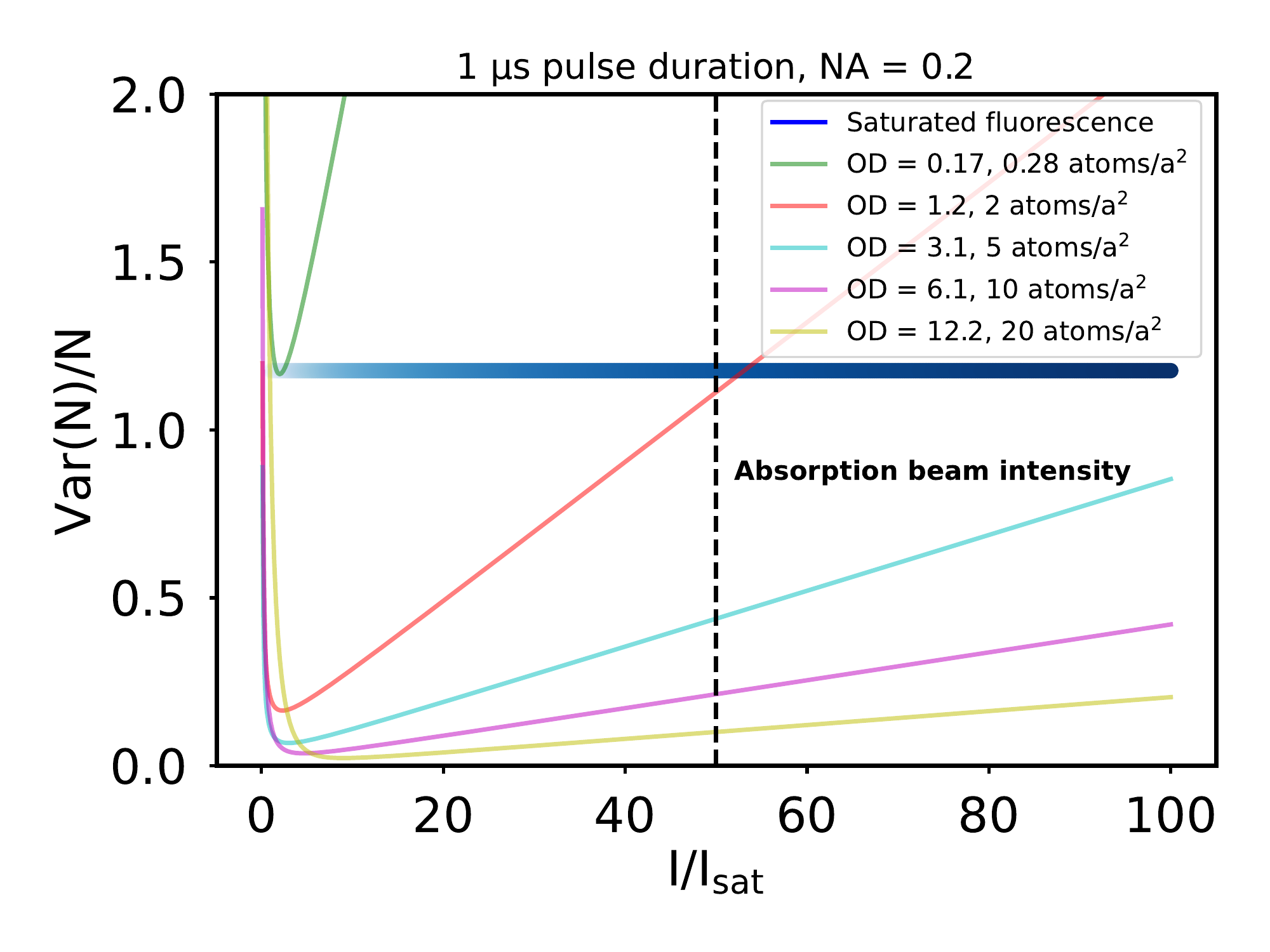}
    \caption{SNR comparison between absorption and fluorescence imaging. The relevant imaging parameters from the main figures of the paper are used for this calculation. For absorption imaging the atom count variance scales inversely proportional with intensity in the non-saturated limit $I \ll I_{sat}$, and proportional with intensity in the high saturation limit. The variance is for both imaging methods proportional to $1/\tau$. In the fully saturated regime (and assuming no technical noise) the normalized variance for fluorescence imaging is independent of atomic column density. To avoid imaging defects at the high densities used in clock operation, an $I/I_{sat} > 50$ was used in all imaging measurements. The black dashed line indicates the intensity used for our inverse Abel measurements. }
    \label{fig:figS2}
\end{figure}

\subsection{HTSE calculation}

To accurately model the density distribution in our 3D lattice, we use a High Temperature Series Expansion (HTSE) calculation in the atomic limit. The general Hamiltonian for SU(N) symmetric fermions in a 3D lattice in the atomic limit takes the following form:

\begin{equation}H_{AL} = \frac{U}{2} \sum_{i, \sigma \neq \sigma^{'}} \hat{n}_{i, \sigma}\hat{n}_{i, \sigma^{'}} + \sum_{i, \sigma} V_{i}\hat{n}_{i, \sigma}\end{equation}

On a lattice site $i$, there are just two competing energy scales: an interaction energy $U$ between particles and a position dependent energy offset $V_{i}$ according to the harmonic confinement. By using the local density approximation $\mu = V(x, y, z) - \mu_{0} $, where  $V(x, y, z) = \frac{1}{2}m(\omega_x^2 x^2 + \omega_y^2 y^2 + \omega_z^2 z^2)$ and $\mu_{0}$ corresponds to the peak chemical potential in the lattice. For the spin-polarized system in this work, U = 0 and the calculations are substantially simplified. 

Ultimately, we want to express the density distribution $n(\mu, T, r)$ in terms of the chemical potential, atomic temperature, and position in the lattice.  On a lattice site $i$, we express the Grand partition function $\mathcal{Z}$ and Grand potential $\Omega$ :

    \begin{equation}\mathcal{Z}(\mu, T, r) =  \sum_{\sigma = 0}^{N = 1} {{N}\choose{\sigma}} e^{-\beta \mu \sigma }\end{equation}
    
    $$ \Omega = -k_{B} T \textrm{ln}(\mathcal{Z})$$

From here, we determine the entropy and occupancy per lattice site $i$:

$$s(\mu, T, r) = -\frac{\partial \Omega}{\partial T} = k_{B}\:\textrm{ln}(\mathcal{Z}) + \Delta s$$


\begin{equation}\Delta s = \frac{ k_{B}}{\mathcal{Z}}  \beta \mu e^{-\beta \mu }\end{equation}

 \begin{equation}
n(\mu, T, r)  = -\frac{\partial \Omega}{\partial \mu} = \frac{1}{Z(\mu, T)}  e^{-\beta  \mu}\end{equation}

We accurately determine the total atom number $N_{lat}$ from in situ absorption imaging and total entropy $S_{lat}$ via time-of-flight fitting to a non-interacting Fermi-Dirac profile.  Similarly, we express the entropy $s$ and occupation $n$ on a given lattice site using Eq. 8 and Eq. 9 expressed in terms of $T$ and $\mu$.  We then determine global fitting parameters $T$ and $\mu_{0}$ to ensure the integrated entropy and occupancy over all lattice sites equals our experimentally measured values. After determining $\mu_{0}$ and $T$ to realize the equality in Eq. 9, we calculate $n(\mu, T, r)$. A linecut of $n(\mu, T, r)$ at $z = 0$ is plotted in Fig.~4(b).

\subsection{Inverse Abel transform}

We outline our reconstruction procedure here using measurements of the atomic cloud aspect ratios and an inverse Abel transform: First, we use saturated absorption images along a vertical axis aligned with $z$ and a horizontal axis aligned with $x$ corresponding to Fig.~1(a) and Fig.~1(b) to determine  the aspect ratios $\omega_{x}/\omega_{y}$ and $\omega_{x}/\omega_{z}$ respectively. Next, we perform an inverse Abel transform on the Fig.~1(a) image to reconstruct an initial three-dimensional density distribution. Given there is no axis of cylindrical symmetry in our system geometry, the reconstructed density from the inverse Abel transform must be appropriately re-scaled. 

Treating our system as an ellipsoid with radii $r_{x}$, $r_{y}$, $r_{z}$ and $N$ atoms the density is $n_{lat} = N/V_{lat}$ where $V_{lat} = \frac{4}{3} \pi r_{x} r_{y} r_{z}$. We extract the inverse Abel transform for the Fig.~1(a) image along the $x$ axis, given the largest Band insulator plateau will occur along the axis with the weakest harmonic confinement. The density distribution from this procedure assumes a volume of $V_{Abel} = \frac{4}{3} \pi r_{x} r_{x} r_{y}$. Thus we scale the extracted density by $n_{Abel}$/$n_{lat}$ = $\frac{r_{z} }{r_{x}}$ = $\omega_{z}/\omega_{x}$ using the measured aspect ratio from Fig.~1(b).  Given excess noise around the origin, the $x$ = $0$ point is interpolated with the neighboring point in Fig.~4(a).  This reconstruction procedure was cross-checked with simulated density distributions to ensure its fidelity. The three-point Abel transform method was used for this work, which has been independently studied to verify its fidelity \cite{hickstein2019direct}. 

\subsection{QPN calculation}

To calibrate our atom number, we analyze quantum projection fluctuations using the narrow-linewidth clock transition between the $^{1}S_{0}$ and $^{3}P_{0}$ states in $^{87}\textrm{Sr}$. Using a clock laser stabilized to our 8 mHz linewidth silicon reference cavity, rotation noise due to laser instability can be neglected in these measurements \cite{matei20171}.  Additionally, fluctuations in total counts are $<$ $2 \%$ and not a limiting systematic for determining the atom number calibration. Referenced in many texts \cite{itano1993quantum}, by preparing atoms in a superposition of $^{1}S_{0}$ to $^{3}P_{0}$ the variance $\mathbb{V}$ of the measured excitation fraction is related to the mean atom number $\bar{N}$ and mean excitation $\bar{p}_{e}$ by:

\begin{equation}
    \mathbb{V}_{QPN} = \frac{\bar{p}_{e}(1 - \bar{p}_{e})}{\bar{N}} 
\end{equation}

To determine this variance, we do many subsequent measurements of $p_{e}$ under identical operating conditions. For a measurement $i$ to determine $p_{e}^{i}$, two fluorescence counts $\tilde{C}_{g}^{i}$ and $\tilde{C}_{e}^{i}$ are read off a region of interest of our camera including our atoms. These counts are subtracted by two averaged dark frames $\bar{B}_{g}$ and $\bar{B}_{e}$ to yield $C_{g}^{i} = \tilde{C}_{g}^{i} - \bar{B}_{g}$,  $C_{e}^{i} = \tilde{C}_{e}^{i} - \bar{B}_{e}$. We would like to determine the coefficient $a$ that satisfies $N_{e}^{i} = a C_{e}^{i}/ \tau$, $N_{g}^{i} = a C_{g}^{i} / \tau$. We can immediately see that the excitation fraction has no dependence on this coefficient:

\begin{equation}
    p_{e}^{i} = \frac{\cancel{a}C_{e}^{i}  }{\cancel{a}C_{e}^{i}  + \cancel{a}C_{g}^{i}}
\end{equation}

However, the total atom number $N^{i} = a(C_{e}^{i}  +C_{g}^{i})/\tau = a C_{t}^{i}/\tau$ does. Rewriting Eq. 10, we see a measurement of the variance $\mathbb{V}_{QPN}$, the mean excitation $\bar{p}_{e}$, and the mean total counts $\bar{C}_{t}$ can determine $a$. 

\begin{equation}
    \mathbb{V}_{QPN} = \frac{\bar{p}_{e}(1 - \bar{p}_{e})}{a \bar{C_{t}}/\tau }
\end{equation}

The coefficient $a$ can be interpreted as the "atoms per count per pulse duration". In principle, with knowledge of the quantum efficiency, gain, scattering rate, numerical aperture, and radiation pattern one could calculate this value. Practically, assumptions about the radiation pattern based on the quantization axis and probe light polarization make this calculation more difficult. In practice, it is much more straightforward to directly measure $a$ than to individually measure each of these values with high accuracy.  

The observed variance of the excitation fraction $\mathbb{V}_{p_{e}}$ has contributions from quantum projection noise (QPN), photon shot noise (PSN), and camera readout noise (RN): 

\begin{equation}
    \mathbb{V}_{p_{e}} = \mathbb{V}_{QPN} + \mathbb{V}_{PSN} + \mathbb{V}_{RN}
\end{equation}

Here $g$ is the detector gain in units of counts per electron.

\begin{equation}\mathbb{V}_{PSN}  = \frac{\bar{p}_{e}(1 - \bar{p}_{e})}{\bar{C_{t}}} \times g \end{equation}

\begin{equation}\mathbb{V}_{RN} = \frac{\mathcal{R}^{2}}{\bar{C_{t}}^{2}} ( 2 \bar{p}_{e}^{2} - 2 \bar{p}_{e}  + 1)\end{equation}

  $\mathbb{V}_{PSN}$ can be understood intuitively considering the ratio $\mathbb{V}_{QPN}/\mathbb{V}_{PSN}$. The number of signal electrons (equivalently the number of collected photons multiplied by the camera quantum efficiency) per atom determines the relative scaling of $\mathbb{V}_{QPN}$ and $\mathbb{V}_{PSN}$. 

\begin{equation}
    \frac{\mathbb{V}_{QPN}}{\mathbb{V}_{PSN}} = \frac{1}{g \times a}
\end{equation}

\renewcommand{\thefigure}{S3}
\begin{figure}[htp]
    \centering
    \includegraphics[width=8.5 cm]{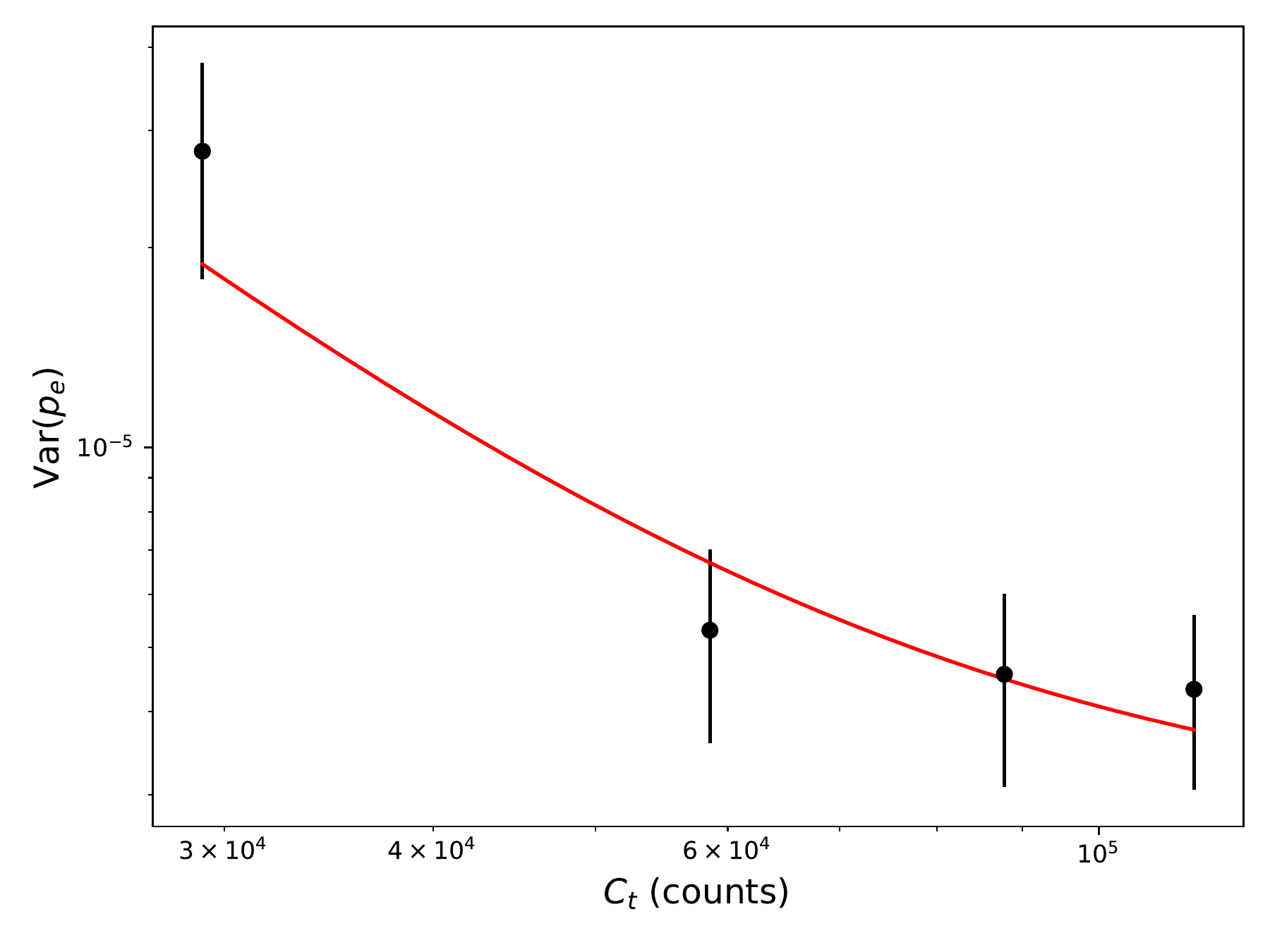}
    \caption{Readout noise calibration. A $\pi$ pulse on our optical clock transition is used so $p_{e}$ $\approx$ 1 and $\mathbb{V}_{p_{e}} = \frac{\mathcal{R}^2}{\bar{C_{t}}^{2}} + C$. We use 4 pulse durations between $5$ and $20$ $\mu$s to vary $C_{t}$. We fit $\mathcal{R} = 100.2 \pm 24.6$ and $C = 2.73 \times 10^{-6} \pm 1.02 \times 10^{-6}$. }
    \label{fig:S3}
\end{figure}

To determine $a$ we need to accurately calibrate $\mathbb{V}_{RN}$ and $\mathbb{V}_{PSN}$. We see at $p_{e} = 1$, $\mathbb{V}_{PSN}$, $\mathbb{V}_{QPN}$ = 0. Thus, measuring $\mathbb{V}_{p_{e}}$ at $p_{e} = 1$ will independently determine $\mathbb{V}_{RN}$. 

We wish to fit $\mathcal{R}$ and ensure it is consistent with the cameras specified readout noise. To extract this value, we use 4 pulse durations between $5$ and $20$ $\mu$s to vary $C_{t}$. This is illustrated in Fig.~\ref{fig:S3}. In practice, we fit

\begin{equation}
    \mathbb{V}_{p_{e}} = \frac{\mathcal{R}^2}{\bar{C_{t}}^{2}} + C
\end{equation}

We fit $\mathcal{R} = 100.2 \pm 24.6$ and $C = 2.73 \times 10^{-6} \pm 1.02 \times 10^{-6}$. For our circular ROI there are $X = 889$ pixels in the masked radius. For the calibrated gain $g = 1.59$  counts/e- and readout noise $r = 2.4$ e- respectively , $\mathcal{R}_{calc} = \sqrt{Xgr}=  94.7$ in  agreement with $\mathcal{R} = 100.2 \pm 24.6$. We note that the gain and readout noise of the camera are close to specification. Dark counts over our $30$ ms exposure are $<$ $.1$ e- and considered negligible.

Next, we wish to determine $a_{QPN}$. To do so, we perform a second measurement at $p_{e} = 0.5$. The variance of this dataset contains contributions from $\mathbb{V}_{QPN}$, $\mathbb{V}_{PSN}$, and $\mathbb{V}_{RN}$. Using the measured $\mathcal{R}$ value, we subtract the $\mathbb{V}_{RN}$ contribution. Next, we fit the data in Fig.~\ref{fig:S4} to:

\begin{equation}
    \mathbb{V}_{p_{e}} = \frac{0.5 (1 - 0.5)}{a \bar{C_{t}}/\tau } + \frac{0.5(1 - 0.5)}{\bar{C_{t}}} \times g 
\end{equation}

We fit $a_{QPN} = 1.72 \pm 0.16$. This is in reasonable agreement with the calculated value of 1.43 assuming $\Gamma$/2 scattering into $4$ $\pi$ while also accounting for the measured quantum efficiency.

\renewcommand{\thefigure}{S4}
\begin{figure}[h]
    \centering
    \includegraphics[width=8.5 cm]{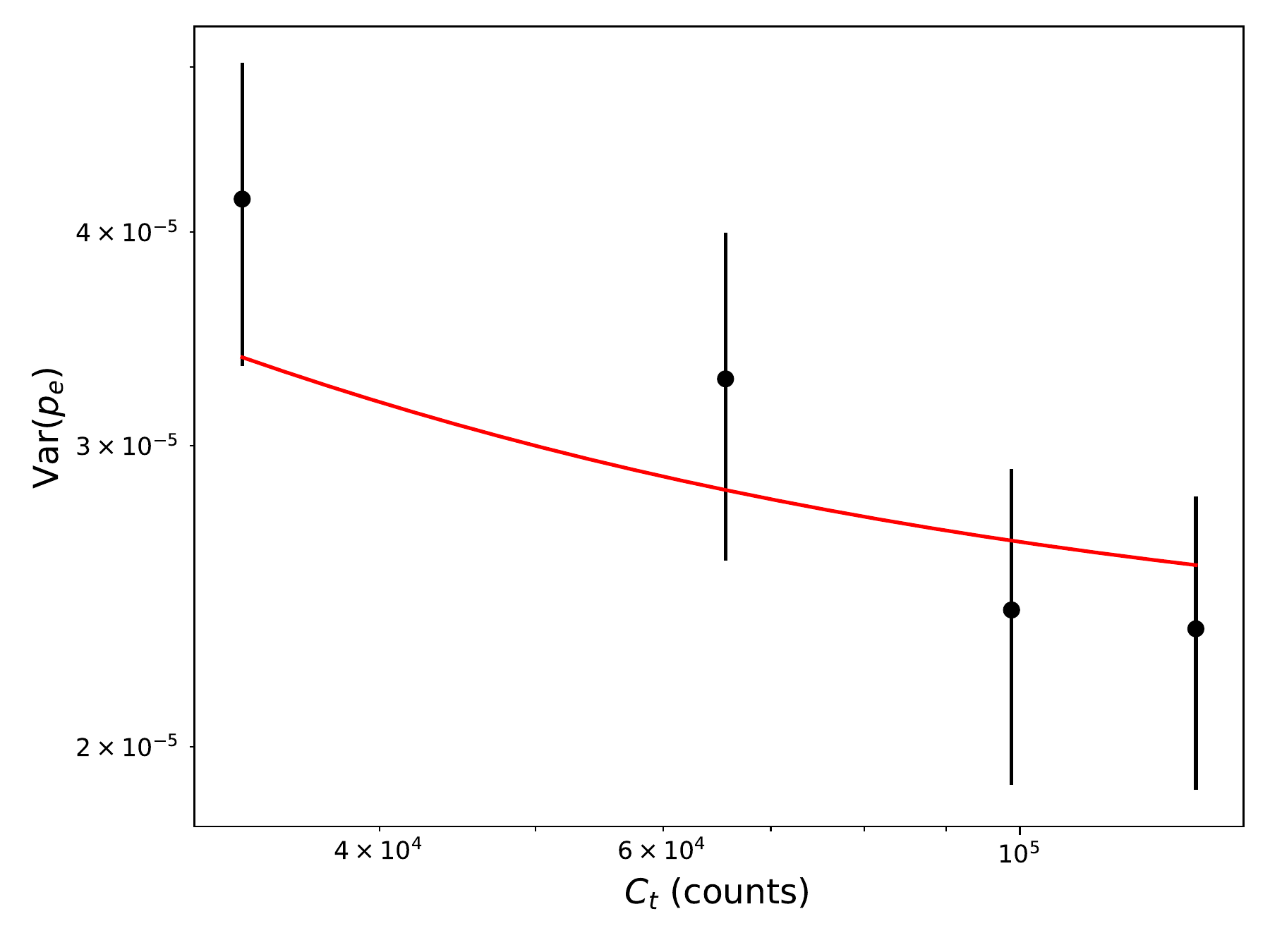}
    \caption{$a_{QPN}$ calibration.  The atoms in our optical lattice are placed in a superposition of the ground and clock states with a  $\pi$/2 pulse so $p_{e}$ $\approx$ 0.5 for these measurements and $\mathbb{V}_{p_{e}}$ is fit to Eq. 18.  We determine $a_{QPN} = 1.72 \pm 0.16$.}
    \label{fig:S4}
\end{figure}

\section{Readout noise}

Here, we derive the readout noise term used in our variance measurements. The expressions used are somewhat different than other literature, given that we use averaged dark frames  $\bar{B}_{e}$ and $\bar{B}_{g}$. Recall, $p_{e} = \frac{C_{e}  }{C_{e}  + C_{g}}$. To determine the readnoise contribution to the excitation fraction, we perform standard error propagation: 

\begin{equation}\mathbb{V}_{RN} = \Big( \frac{\partial p_{e}}{\partial C_{e}} \Big)^2\mathbb{V}(C_{e}) + \Big(\frac{\partial p_{e}}{\partial C_{g}} \Big)^2\mathbb{V}(C_{g})\end{equation}

Here, 
\begin{equation}
\frac{\partial p_{e}}{\partial C_{g}} = \frac{C_{e}}{(C_{e} + C_{g})^{2}} = \frac{p_{e}}{(C_{e} + C_{g})}
\end{equation}

\begin{equation}
\frac{\partial p_{e}}{\partial C_{e}} = \frac{C_{g}}{(C_{e} + C_{g})^{2}} = \frac{1 - p_{e}}{(C_{e} + C_{g})}
\end{equation}

To determine $\mathbb{V}(C_{e})$ consider an $X$ pixel region-of-interest for which we extract $C_{g}$, $C_{e}$ in two separate measurements. Each pixel contains $r$ read noise in electrons. The single pixel read noise in units of counts is thus $g \times r_{i}$. The total noise in this region of interest is summed in quadrature pixel-by-pixel $\mathbb{V}(C_{g}), \mathbb{V}(C_{e}) = \sum_{X}{(r_{i} \times g)^{2}} = Xr^{2}g^{2}$ = $\mathcal{R}^2$. Plugging terms in Eq. 19:

\begin{equation}\mathbb{V}_{RN} = \frac{\mathcal{R}^{2}}{\bar{C_{t}}^{2}} ( 2 \bar{p}_{e}^{2} - 2 \bar{p}_{e}  + 1)\end{equation}


\subsection{Imaging system parameters for Fig.~3(a)}

In Table 1 is a summary of the imaging parameters for the measurements in Fig.~3(a). For Fig.~1 and Fig.~4, a 1 $\mu$s pulse duration was used. In Fig.~3(b), we vary the pulse length between 500 ns and 2 $\mu$s. Atom number fluctuations in time-of-flight absorption imaging for these measurements have a standard deviation less than $2$ $\%$.

\begin{center}
\begin{tabular}{ |p{6cm}|p{2cm}|  }
\hline
\multicolumn{2}{|c|}{Table 1}
\\
\hline
\multicolumn{2}{|c|}{Vertical imaging system} \\
\hline

Numerical aperture & 0.23  \\
Pulse duration & 3 $\mu$s  \\
Total photons scattered per atom at full saturation & 287  \\
Collection efficiency & 1.3 \%  \\
Camera quantum efficiency & 0.85  \\
Imaging system quantum efficiency & 0.65  \\
Calculated photon count per atom & 2.06  \\
Measured photon count per atom & 1.91(1)  \\

\hline
\multicolumn{2}{|c|}{Horizontal imaging system} \\
\hline

Numerical aperture & 0.10  \\
Pulse duration & 3 $\mu$s  \\
Total photons scattered per atom at full saturation & 287  \\
Collection efficiency & 0.25 \%  \\
Camera quantum efficiency & 0.78  \\
Imaging system quantum efficiency & 0.72  \\
Calculated photon count per atom & 0.402  \\
Measured photon count per atom & 0.445(3)  \\

\hline
\end{tabular}
\end{center}

\bibliography{references}